\documentclass{article}

\usepackage{PRIMEarxiv}
\usepackage{amsmath}  
\usepackage[utf8]{inputenc} 
\usepackage[T1]{fontenc}    
\usepackage{hyperref}       
\usepackage{url}            
\usepackage{booktabs}       
\usepackage{amsfonts}       
\usepackage{nicefrac}       
\usepackage{microtype}      
\usepackage{lipsum}
\usepackage{fancyhdr}       
\usepackage{graphicx}       
\graphicspath{{media/}}     
\usepackage{color,soul}
\title{Medical Image De-identification, Cleaning, and Compression Using PyLogik
\thanks{\textit{\underline{Corresponding author}}: Adrienne Kline (askline1@gmail.com)} 
}

\author{
  Adrienne Kline*\\
  Department of Preventative Medicine\\
  Institute for Augmented Intelligence in Medicine\\
  Northwestern University\\
  Chicago, USA\\
  \texttt{askline1@gmail.com} \\
   \And
  Vinesh Appadurai\\
  Division of Cardiology, Department of Medicine,\\
  Northwestern University Feinberg School of Medicine\\
  Bluhm Cardiovascular Institute,\\
  Northwestern Memorial Hospital\\
  Chicago, USA\\
  School of Medicine,\\
  The University of Queensland, Australia\\
  \texttt{vinesh.appadurai@nm.org} \\
  \And
  Yuan Luo\\
  Department of Preventative Medicine\\
  Institute for Augmented Intelligence in Medicine\\
  Northwestern University\\
  Chicago, USA\\
  \texttt{yuan.luo@northwestern.edu} \\
  \And
  Sanjiv Shah\\
  Division of Cardiology, Department of Medicine,\\
  Northwestern University Feinberg School of Medicine\\
  Bluhm Cardiovascular Institute,\\
  Northwestern Memorial Hospital\\
  Chicago, USA\\
  \texttt{sanjiv.shah@northwestern.edu} \\
}

\begin{document}
\maketitle

\begin{abstract}
Leveraging medical record information in the era of big data and machine learning comes with the caveat that data must be cleaned and de-identified. Facilitating data sharing and harmonization for multi-center collaborations are particularly difficult when protected health information (PHI) is contained or embedded in image meta-data. We propose a novel library in the Python framework, called PyLogik, to help alleviate this issue for ultrasound images, which are particularly challenging because of the frequent inclusion of PHI directly on the images. PyLogik processes the image volumes through a series of text detection/extraction, filtering, thresholding, morphological and contour comparisons. This methodology de-identifies the images, reduces file sizes, and prepares image volumes for applications in deep learning and data sharing. To evaluate its effectiveness in processing ultrasound data, a random sample of 50 cardiac ultrasounds (echocardiograms) were processed through PyLogik, and the outputs were compared with the manual segmentations by an expert user. The Dice coefficient of the two approaches achieved an average value of 0.976. Next, an investigation was conducted to ascertain the degree of information compression achieved using the algorithm. Resultant data was found to be on average $\sim72\%$ smaller after processing by PyLogik. Our results suggest that PyLogik is a viable methodology for data cleaning and de-identification, determining ROI, and file compression which will facilitate efficient storage, use, and dissemination of ultrasound data. Variations of the pipeline have also been created for use in other medical imaging data types.
\end{abstract}

\keywords{ultrasound \and de-identification \and data cleaning \and image processing \and machine learning}

\section{Introduction}
Large conglomerate imaging datasets have facilitated open-source repositories of information for development of machine learning and artificial intelligence \cite{ADNI2022} \cite{OAI2022} \cite{COCO2022}. In fact, these databases provide current state-of-the-art benchmarks for the development and testing of novel methodologies \cite{biswas2019state} \cite{hering2022learn2reg}. For example, algorithmically derived knowledge from 2- and 3-dimensional imaging arrays has contributed to health care, sometimes even outperforming human experts in applications such as chest x-rays, retinal scans, and echocardiography assessments \cite{patel2019human} \cite{kamran2019optic} \cite{madani2018fast}. However, gleaning insights from noisy health information is often very difficult, even though it is the foundation on which much work in statistics, machine learning and artificial intelligence rests. Data cleaning is a necessary first step for facilitating inferences drawn from analyses of those data. Cleaning includes outlier detection, removal of incomplete data, imputation, and de-identification – the latter is one focus of the current study. In addition, images, such as ultrasounds, create very large data sets which contain irrelevant as well as relevant information. To address these issues, the aims of the current work are threefold: 1) Provide a framework for detecting and extracting text from medical images, 2) Clean ultrasound images in preparation for sharing or algorithm development and 3) Image truncation to provide more efficient storage space on secondary servers.

Major infrastructures and workflows define how medical images are stored and accessed in a hospital network. Picture Archiving and Communication Systems (PACS) use the Digital Imaging and Communications in Medicine (DICOM) standard, which defines data formats, storage organization and communication protocols for digital medical imaging \cite{gibaud2008dicom}. DICOM files can support multiple types of medical information, such as images, waveforms, structured reports, etc. These structured objects include a header that contains metadata used to represent the DICOM Information Model \cite{graham2005dicom}. For instance, it may include attributes such as the patient’s name, phone number, social security number, the institution’s name, and imaging modality/manufacturer. Additionally, there is a DICOM attribute reserved for pixel data which may contain a single image or multiple frames (e.g., cine-loops). The pixel data may be in a raw format or compressed using a variety of standards, including JPEG, JPEG Lossless, JPEG 2000, MPEG2 and MPEG4.

Due to the risk of unauthorized sharing of protected health information (PHI), efforts have been made to de-identify
medical images for broader applications. Newhauser et al. worked to automate DICOM de-identification using Optical Character Recognition (OCR) to detect and mask all text within the image arrays \cite{newhauser2014anonymization}. Another approach to text masking uses a Convolutional Recurrent Neural Network (CRNN). This architecture is a combination of Convolutional Neural Network (CNN), Recurrent Neural Network (RNN), and Connectionist Temporal Classification (CTC) and used for image-based sequence recognition tasks \cite{shi2016end}. Attributes within the image such as curves, lines, intersections, and loops, and how they occur in sequence relative to one another at different levels of analysis provide de-identified image data. Previous work in this area has included that by Fezai et al. who applied a convolution neural network architecture using a simple auto-encoder to classify magnetic resonance imaging (MRI) equipment \cite{fezai2022deep} from the metadata contained in the image array. Monteiro et al. demonstrated a convolution neural network (CNN) pipeline for de-identification of ultrasound DICOM images that removes identifying text only, while leaving non-identifying text intact \cite{monteiro2017identification}. Their software is available as a software-as-a-service, with an anonymization success of 89.2\%. Huang et al. showcased a method following HIPAA privacy rules which also implemented and OCR \cite{huang2009privacy} with a recorded success rate of 65\%. Their algorithm, however, struggled to distinguish characters such as "Bg", from "B9" or "B1 from "Bl" and "B0 or "Bo". Oftentimes this is a difficult task due to nuisance features arising from medical image background, which complicates the OCR methodology. However, completely de-identifying images is not practical if they are going to be used for secondary applications \cite{parker2021canadian}. Therefore, the aim of de-identification should be to retain meaningful metadata (e.g., age, race, sex) information to allow for translation applications and determine the generalizability of the algorithm(s) developed.

Building on the methodologies and limitations of previous work, we also propose the use of an OCR for text detection. However, unlike some of the previous work, we hypothesize that the removal and sequestering all text to a separate file structure provides a two-fold advantage. First, by removing all text, we increase our assurances that PHI is, in fact, removed. It also places relevant patient information in a format that is easily accessible to end users. Second, extracting ’burned-in’ redundant information (as it is repeated in every frame) from the image array permits the truncation of the image information that is now of pertinent interest. Further isolation of the ROI permits truncation of the image, thereby reducing file sizes.

Since large repositories of image data are held on servers for processing, reducing file size is desirable both for
permanent (hard drive) and temporary (RAM) memory storage of the files. In both cases, file size dictates how many images can be held at any given time. Using PyLogik, the file size of each ultrasound image is expected to decrease dramatically, enhancing the storage, and sharing capability of ultrasound images for use in machine learning studies.

\section{Methods}

\subsection{Ultrasound Echocardiogram Data Set}

Ultrasound images typically resemble one of the three formats outlined in Figure 1. The shape differences are by-products of the manufacturing system and the ultrasound probe being used - phased array, curvilinear and linear in Figure 1a, 1b, and 1c, respectively \cite{USprobetypes}. Data were processed as DICOM formatted images; a typical format used in hospital-based PACS \cite{gibaud2008dicom}. These typically contain metadata about the patient, study ID, the machine used to acquire the data, etc \cite{robinson2014beyond}. Reading these image volumes into Python 3 we can separate the metadata from the image arrays and process them as needed. Cardiac ultrasound (echocardiographic) images were acquired from the Northwestern University Echocardiography Core Laboratory (NUECL) repository, this study was approved by the institutional review board (STU00217900). All echocardiograms were done using an institutional review board-approved protocol, and all patients provided written informed consent. A Vivid T8 ultrasound system (GE Healthcare, Waukesha, WI) was used to perform all echocardiographic studies. 

\begin{figure}[h]
  \centering
  \includegraphics[scale=0.6]{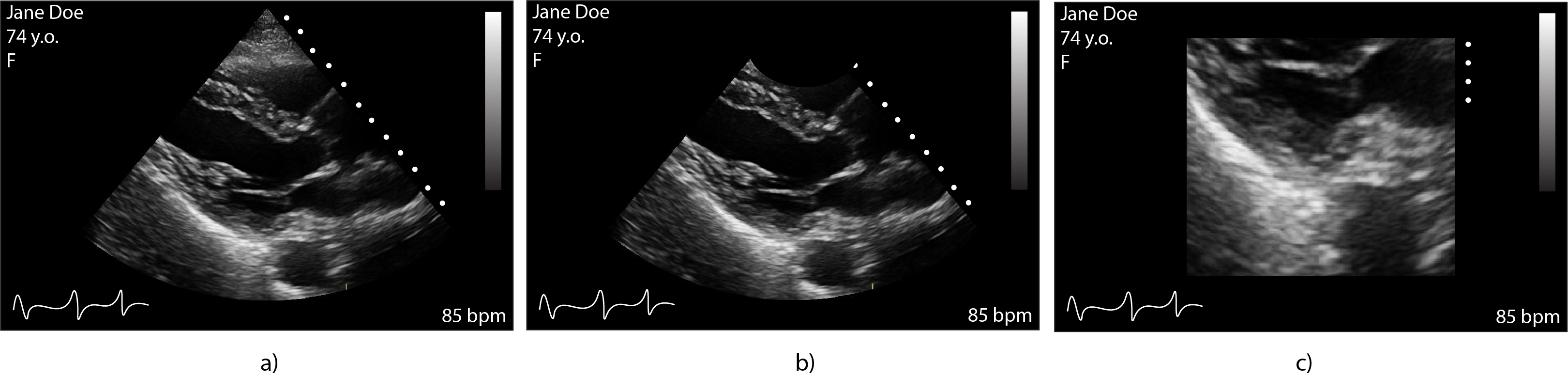}
  \caption{Examples of the structure of a given ultrasound based on probe type. a) phased array, b) curvilinear and c) linear}
  \label{fig:fig1}
\end{figure}

\subsection{Text Masking}

An overview of the workflow is presented in Figure 2. The first step in the pipeline is to detect, extract and mask text-based data contained within the arrays. The text detection framework is implemented using Pytorch \cite{PYTORCH}. The text recognition model is based on a trained convolutional recurrent neural network (CRNN) proposed by Shi et al. \cite{shi2016end} which was trained on datasets IC13 \cite{karatzas2013icdar}, IIIT5k \cite{mishra2012scene} and SVT \cite{wang2011end}. It consists of 3 main components: a) feature extraction, performed using ResNet (a convolutional neural network), and visual geometry group (VGG) neural network b) sequence labeling, using a Long Short Term Memory network (LSTM) and c) decoding that is performed by Connectionist Temporal Classification (CTC) 
 \cite{graves2006connectionist}.

The convolutional layers in VGG use a very small receptive field (3x3, the smallest possible size that still captures left/right and up/down). There are also 1x1 convolution filters which act as a linear transformation of the input, which is followed by a ReLU unit. The convolution stride is fixed to 1 pixel so that the spatial resolution is preserved after convolution. The VGG has three fully-connected layers: the first two have 4096 channels each and the third has 1000 channels, 1 for each class. All VGG’s hidden layers use ReLU (an innovation from AlexNet that has been proven to cut training time). VGG does not generally use Local Response Normalization (LRN), as LRN increases memory consumption and training time with no discernible increase in accuracy \cite{baek2019character}.

A deep bidirectional Recurrent Neural Network (RNN) is built on the top of ResNet. The recurrent layers predict a label distribution $y_t$ for each frame $x_t$ in the feature sequence $x = x_1, ..., x_T$. There are several advantages of leveraging recurrent layers. First, RNN has a strong capability of capturing contextual information within a sequence. The sequence here being sequences within the same image. Using these contextual cues for image-based sequence recognition is more robust than treating each symbol independently. For example, in text detection, certain characters may require several successive kernels to fully encapsulate the word. A long short term memory network LSTM is a type of RNN unit that is specifically designed to address this problem. Thus, the ambiguity that would otherwise exist in individual characters/kernels are more easily distinguished when observing their contexts - i.e. it is easier to recognize "nm" by contrasting the character widths than by recognizing each of them separately. Further, an RNN can back-propagate error differentials back to its input (ResNet), allowing us to jointly train the recurrent and the convolutional layers in a unified network. Lastly, RNN is able to operate on sequences of arbitrary lengths, traversing from starts to end. The structure of an LSTM comprises a memory cell and three gates that perform multiplicative operations: the input, output, and forget gates. The memory cell serves as a container for previous contexts, while the input and output gates enable the cell to retain information over an extended duration. Additionally, the forget gate facilitates the erasure of memory from the cell. Due to its unique architecture, an LSTM can effectively capture long-term dependencies, which frequently arise in sequences of image data.

The CTC enables OCR to be carried out, permitting the training of text recognition models using image and ground truth text pairs. The neural network output matrix encodes text by creating paths that consist of one character per time-step. As such, 'ab' or 'aa' represent conceivable paths \cite{liao2022real}. In a CTC, the neural network takes an input image and outputs a matrix that encodes the probability of each character at each time-step. The CTC algorithm then maps these probabilities to the actual text in the image. This mapping is performed using a dynamic programming approach, which considers all possible alignments between the output probabilities and the target text. During training, the CTC loss function is used to optimize the parameters of the neural network. The CTC loss function calculates the negative log-likelihood of the correct alignment between the output probabilities and the target text, taking into account all possible alignments. The CTC method allows training of OCR systems using pairs of images and ground truth texts. The ground truth texts are used to calculate the CTC loss function during training, which allows the neural network to learn to output the correct text given an input image.

We leverage the conditional probabilities using a CTC layer proposed by Graves et al. \cite{graves2006connectionist}. The probability is defined for label sequence $l$ conditioned on the per-frame predictions $y = y_1,..,y_T$, and it ignores the position where each label in $l$ is located. Consequently, using the negative log-likelihood of this probability as the objective function to train the network, requires only images and their corresponding label sequences, avoiding the labor of labeling positions of individual characters. This is seen in equation 1, where the training set is: $\chi = \{I_i,l_i\} $ $I_i$ is the training image and $l_i$ is the ground truth label.

\begin{equation}
    \mathcal{O} = -\sum_{I_i,l_i \in \chi} \textrm{log} p(l_i|y_i)
\end{equation}

To calculate conditional probability, the input is a sequence $y = y_1, ..., y_T$ where $T$. Here, each $y_t$ in $R^{|L'|}$ is a probability distribution over the set $L^{'} = L \cup \{blank\}$, where $L$ contains all labels in the task, as well as a 'blank' label denoted by " ". A sequence-to-sequence mapping function $B$ is defined on sequence $\pi$ in $L'^{T}$, where $T$ is the length of the sequence. $B$ maps $\pi$ onto $l$ by first removing repeated labels and the blanks. For example, $B$ maps "--jj-a-a-nn-ee--" (where '-' represents blank spaces) onto "jane". The conditional probability is defined as the sum of probabilities of all $\pi$ that are mapped by $B$ onto $l$ (Eq. 2). Using a lexicon free transcription, the sequence $l^*$ can be approximated as $l^* \approx B(\textrm{argmax}_\pi p(\pi|y))$.

\begin{equation}
    p(l|y) = \sum_{\pi:\mathcal{B}(\pi)=l} (\pi|y)
\end{equation}

The smallest possible bounding box of the area identified to contain each path of text is replaced with matrices with an intensity value of 0, thus masking and effectively destroying ’burned-in’ image metadata. Each unique ’path’ of text is extracted and written to a respective comma separated value file, later available to the user to re-associate with the image on an as-needed basis.

\subsection{Determining Region of Interest (ROI)}

We exploit that there exist frame-to-frame intensity differences amongst the images to ascertain the ROI. Images
undergo Gaussian smoothing using an adaptive kernel that adjusts with respect to array size (Eq. 3), a thresholding is performed based on the intensity of the background in the image. This is set to a default value, but can be changed by the user as needed and highlighted below. Following smoothing and thresholding, morphological image operations are performed (Eq. 4) and area filters are applied.

\begin{equation}
    G_\sigma(x,y) = \frac{1}{2\pi\sigma^2}\exp(-\frac{x^2 + y^2}{2\sigma^2})
\end{equation}

\begin{equation}
    A \bullet B = (A \oplus B) \ominus B
\end{equation}

Following classic image processing techniques, further geometric assessments are made based to determine the best complete ROI to capture. Three points on the previously established ROI are used to complete this task. These three points represent either points on the circumference of the circle (Figure 3a and 3b) or points along the lower portion of a rectangular ROI (Fig. 3c), denoted with a green $X$. The slope between these points is calculated. It is this first step that allows determination of a rectangular or circular ROI, as the expected slope in a rectangular ROI will be approximately 0. The midpoints of the established chords (in the case of a circle) are calculated, and the negative reciprocal of the slope is used to determine the perpendicular slope of the line (Eq. 5). Where these two perpendicular lines (dashed orange) in Figure 3 intersect will present the center $C$ of the circle. Again, non-convergence of the perpendiculars or non-convergence within a specified space suggests that the probe type is linear, and therefore the ROI is rectangular in nature. This can be seen in Fig. 3. Knowing $C$, we can determine the radius $r$ of the circle of where the radius is the argmax of the radius calculated for each of the 3 original points $X$ to preserve a maximal ROI. The angle $\theta$ subtended by the left and right-most radii is calculated (Eq. 6) to create a pie-shaped wedge mask. Of note, equation 6 will always find the acute angle. To differentiate structures seen in Figure 3a and 3b, we can exploit that in 3a, $C$ is well approximated near or within the pre-specified ROI, otherwise a secondary circle is calculated using available information. The secondary circle with radius $r_2$ is used to create the notch seen at the top of the wedge in Figure 3b.

\begin{equation}
    (y-y_1) = m(x-x_1)
\end{equation}

\begin{equation}
    \theta = \arctan(\frac{m_1 - m_2}{1 - m_1*m_2})
\end{equation}

A final check is performed to ensure that the ROI established from classic image processing technique is a reasonable subset of the final ROI ($ROI_{im}$ $\subseteq$ $ROI_{final}$). If this is not achieved, the system will default to using only morphologically identified ROI.

\begin{figure*}[h]
  \centering
  \includegraphics[scale=0.81]{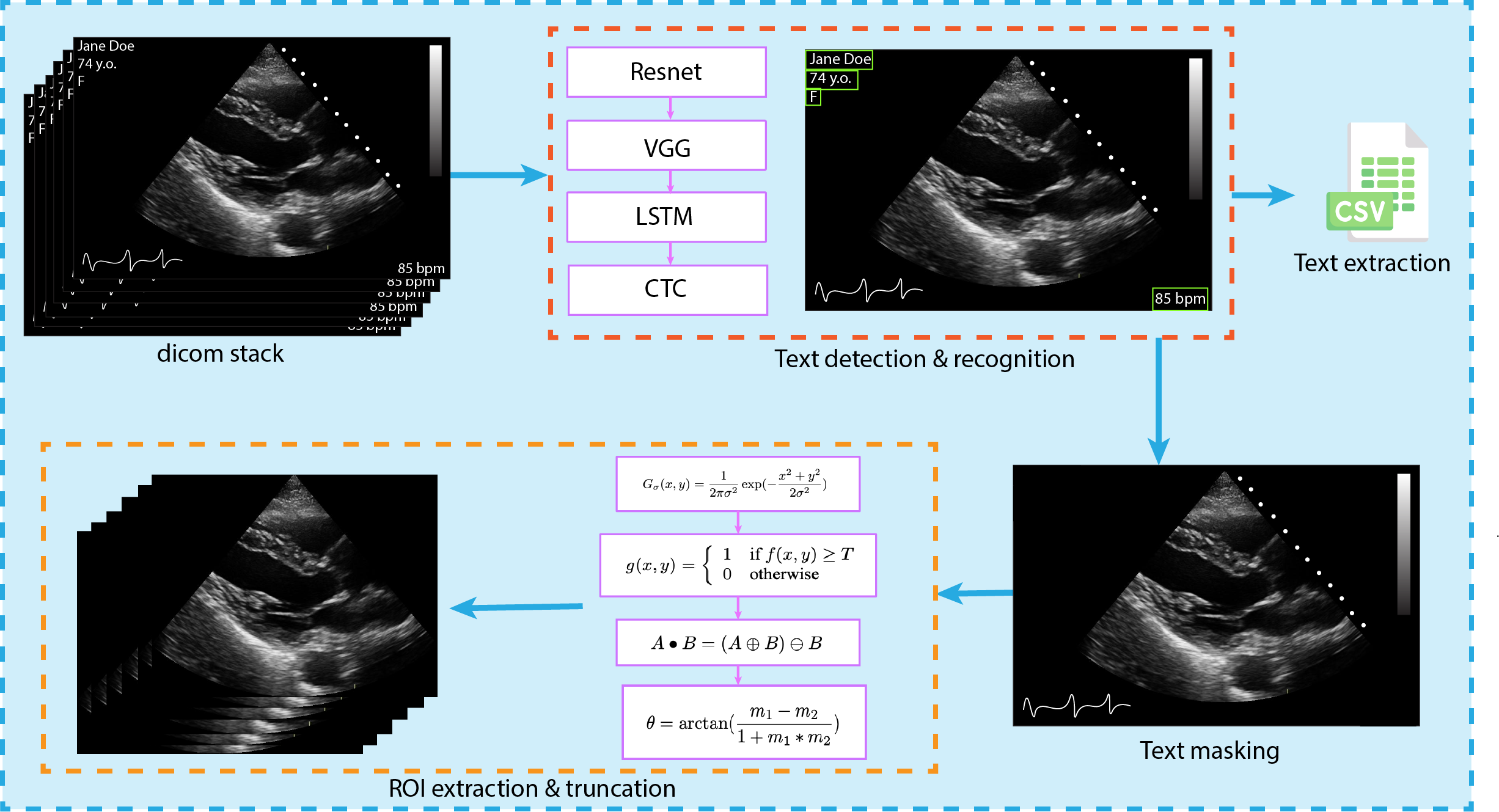}
  \caption{Overview of Data Cleaning and Anonymization. Data is loaded (DICOM, JPEG stack, or video file) and undergoes text detection and removal by implementation of an OCR and masking procedure. All detected text is written to a separate file. Images are further refined by removing nuisance elements in the image. Isolation of the ROI is performed using a series of filtering, morphological and geometric operations.}
  \label{fig:fig1}
\end{figure*}

\begin{figure}[h]
  \centering
  \includegraphics[scale=0.7]{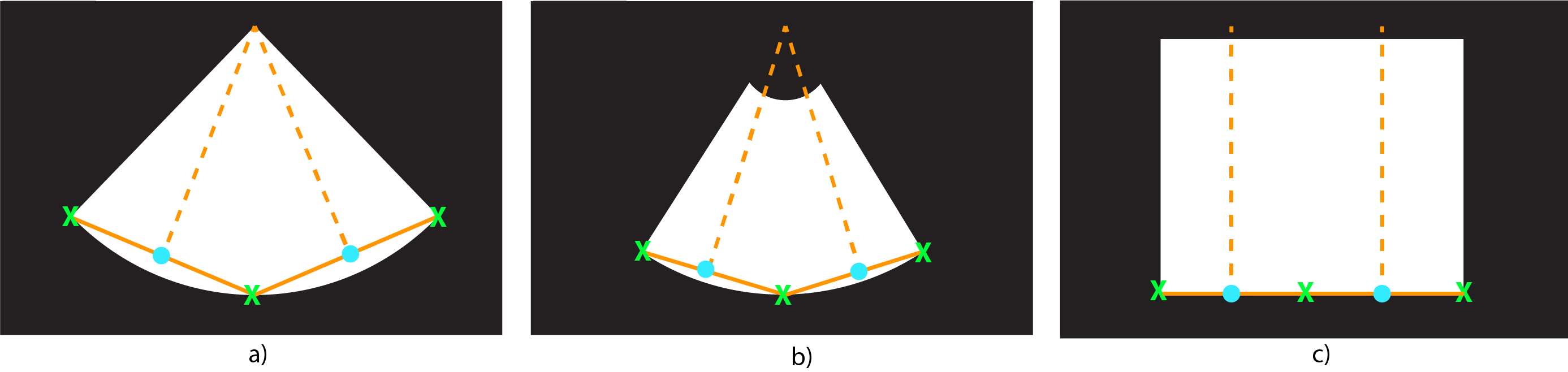}
  \caption{Geometric shape determination. a) pie-shaped wedge (phased array probe) b) notched wedge (curvilinear probe) c) rectangular (linear probe)}
  \label{fig:fig2} 
\end{figure}

\newpage

\subsection{Use Case Example}
We demonstrate the utility of the work and how end users can go about implementing these algorithms in their respective applications. Our software can be installed via terminal by the command: \texttt{\small{pip install pylogik}}. Pieces of the pipeline can be used jointly or in isolation, and to this end we have created several different functions. Compatible image types include 2D (grayscale), 3D (grayscale multiple frames or 3 channel RGB) and 4D (multiple frames and RGB information) with readability for dicom, png, jpg, jpeg and NIfTi image types. Additional files such as files that are skipped and processed are recorded in log files in the destination folder. For our analysis pertaining to the echocardiograms, the function \texttt{\small{deid\_us}} was used. Functions can be imported and used as follows:

\texttt{\small{from pylogik.image import deid}}\\
\texttt{\small{from pylogik.image import im\_analysis}}

\begin{enumerate}
\item Instantiate the \texttt{deid} function from PyLogik. This only removes burned in text from the image, writes it to a .csv file, with the file name (for crosswalk purposes) and writes image frame(s) to lossless JPEG(s):\\
\texttt{\small{deid.deid(input\_path = "path\_to\_files", output\_path="path\_to\_save\_files", rename\_files = False, threshold = 0)}}\\
\textit{input\_path} : path to image files\\
\textit{output\_path} : path to save new image files and .csv text files\\
\textit{rename\_files} : False (default) changes filename to a series of 10 randomly selected alphanumerics\\

\item Instantiate the \texttt{deid\_clean} function from PyLogik. This removes burned in text from the image, writes it to a .csv file, with the file name (for crosswalk purposes) and writes image frame(s) to lossless JPEG(s)(removes/extracts text and removes small scale features):\\
\texttt{\small{deid.deid\_clean((input\_path = "path\_to\_files", output\_path="path\_to\_save\_files", rename\_files=False, threshold = 0)}}\\
\textit{input\_path} : path to image files\\
\textit{output\_path} : path to save new image files and .csv text files\\
\textit{rename\_files} : False (default) changes filename to a series of 10 randomly selected alphanumerics\\
\textit{threshold} : 0 (default)\\

\item Instantiate the \texttt{deid\_one} function from PyLogik. This removes burned in text from the image, writes it to a .csv file (with the file name for crosswalk purposes), keeps the single most salient item in the image - compresses accordingly, and writes image frame(s) to lossless JPEG(s) \\
\texttt{\small{deid.deid\_one(input\_path = "path\_to\_files", output\_path="path\_to\_file\_save", rename\_files=False, threshold = 0)}}\\
\textit{input\_path} : path to image files\\
\textit{output\_path} : path to save new image files and .csv text files\\
\textit{rename\_files} : False (default) changes filename to a series of 10 randomly selected alphanumerics\\
\textit{threshold} : 0 (default)\\

\item Instantiate the \texttt{deid\_us} function from PyLogik. This removes burned in text from the image, writes it to a .csv file (with the file name for crosswalk purposes), processes and compresses images according to methods outlined above, and writes image frame(s) to lossless JPEG(s) :\\
\texttt{\small{deid.deid\_us(input\_path = "path\_to\_files", output\_path="path\_to\_file\_save", rename\_files=False, thresh=0)}}\\
\textit{input\_path} : path to image files (DICOM, JPEG, or video)\\
\textit{output\_path} : path to save new image files and .csv text files\\
\textit{thresh} : integer value of the threshold in the image (default = 0). If unclear to user, can use default or use color\_select tool to capture background intensity from sample image\\
\textit{rename\_files} : False (default) changes filename to a series of 10 randomly selected alphanumerics\\
\textit{threshold} : 0 (default)\\

\item Instantiate the \texttt{find\_txt} function from PyLogik. This only finds text in the image, writes it to a .csv file to the specified output folder, it does not write images\\
\texttt{\small{deid.find\_txt(input\_path = "path\_to\_files", output\_path="path\_to\_save\_files")}}\\
\textit{input\_path} : path to image files\\
\textit{output\_path} : path to save .csv files\\

\item Instantiate the \texttt{dicom\_2\_jpg} function from PyLogik. This reads in a stack of images and reads them back out as a stack of JPEGs with possibility to rename files:\\
\texttt{\small{deid.dicom\_2\_jpg(input\_path = "path\_to\_files", output\_path="path\_to\_file\_save", rename\_files=False)}}\\
\textit{input\_path} : path to image files\\
\textit{output\_path} : path to save new image files and .csv text files\\
\textit{rename\_files} : False (default) changes filename to a series of 10 randomly selected alphanumerics\\

\item Other functions available to the user:\\
Dice score calculation
\texttt{\small{im\_analysis.dice\_score(pred, true, k=1)}}\\
\textit{pred} - array of the predicted segmentation\\
\textit{true} - array of the ground truth segmentation\\
\textit{k} - value to perform matching on (default = 1)\\
Returns: dice score (float)\\

\texttt{\small{im\_analysis.imshowpair(pred, true, color1 = (124,252,0), color2 = (255,0,252), show\_fig=True)}}\\
visualization of dice calculation
\textit{pred} - array of the predicted segmentation\\
\textit{true} - array of the ground truth segmentation\\
\textit{color1} - first color to show unique values from first image\\
\textit{color2} - second color to show unique values from second image\\
Returns: array and graphical plot\\

\texttt{\small{im\_analysis.color\_select(img)}}\\
Allows the user to select a point from an image used in color determination
\textit{img} - 2D image array \\
Returns: Tuple representing image color returned\\
\end{enumerate}

\subsection{Algorithm-Expert Comparison}
To ascertain the efficacy of our algorithm, we had an expert user (an experienced cardiologist with expertise in cardiac ultrasound) perform manual segmentation on a randomly selected subsample of n=50 from the aforementioned NUECL echocardiography repository. Dice coefficients were calculated (Eq. 6) to ascertain the degree of intersection between image sets relative to the total area identified by each procedure alone \cite{zou2004statistical}. Our domain expert identified the ROI by interacting with each image and clicking on boundary points, as seen in Figure 4. A manual review was also performed on 200 different randomly selected echocardiograms, checking outcomes from the software to ensure no patient related data leaked through.

\begin{figure}[h]
  \centering
  \includegraphics[scale=0.65]{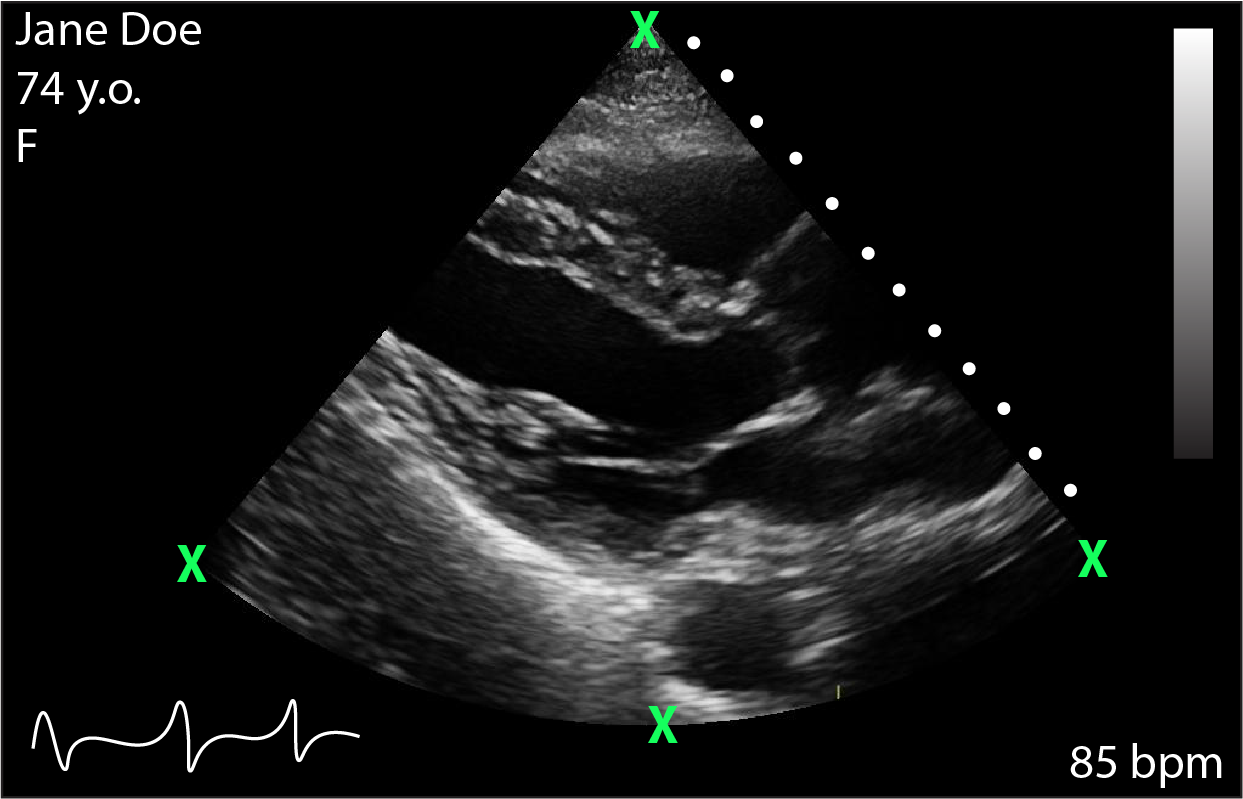}
  \caption{Boundary detection of ROI as defined by user expert}
  \label{fig:fig2} 
\end{figure}

\begin{equation}
    DSC = \frac{2*|X\cap Y|}{|X|+ |Y|}
\end{equation}

\subsection{File Size}
To compare the image size of the ultrasound files, we used the same random sample of 50 echocardiograms as was used in the masking and ROI analyses. We designate the initial size of the files as 100\% whereby the average improvement was quantified in both absolute terms and relativistic ones.

\section{Results}
\subsection{De-identification}
The manual review, performed on 200 different echocardiograms, found no residual PHI or text contained in the image. Further, the text detected and output to .csv files accurately represented ground truth text within the images.

\subsection{Comparison with Expert Annotation}
An average Dice score of 0.976 was achieved. Differences between automated and manual approaches predominantly occurred at boundary regions. Three examples comparing the agreement between the automated and manual segmentation of ROI of the three morphological structures are shown in Figure 5. The white portion of the mask denotes areas of overlap between the two modalities, X $\cap$ Y; green where X $\not\subset$ Y and magenta where Y $\not\subset$ X, where X and Y are the result of automated and manual segmentation procedures, respectively. Further, the manual review of the 200 echocardiograms was successful in that no patient data or text had leaked through into the final image arrays.

\begin{figure}[h]
  \centering
  \includegraphics[scale=0.6]{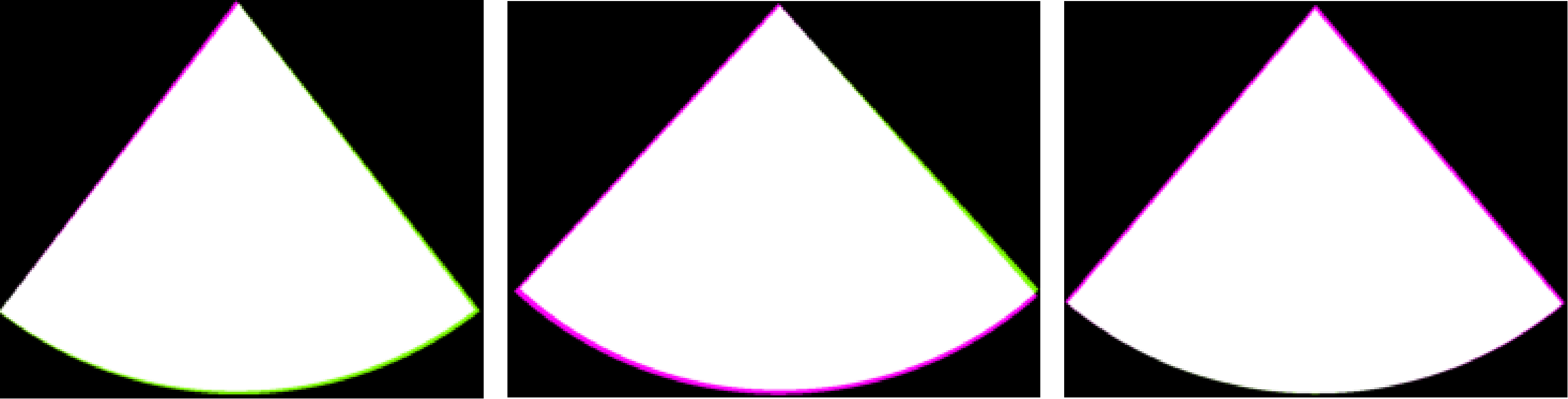}
  \caption{Graphical representation of dice score between expert annotation and algorithm. Graphical representation is implemented using .imshowpair function from the PyLogik library}
  \label{fig:fig2} 
\end{figure}

\subsection{Information Compression}
The results of the file size comparison analysis are reported in Table 1. As we can see, the metadata became nearly negligible (209 KB) with respect to the image file size ($\sim$274 MB). The compression of the image samples neared 72\%.

\begin{table}[h]
 \caption{File Size Before and After Algorithm}
  \centering
  \begin{tabular}{lll | l | l}
    \toprule
    
    \multicolumn{5}{c}{After}                   \\
    \cmidrule(r){2-4}
    Before    & Image Data  & MetaData & Total & Compression\\
    \midrule
    969 MB (100\%) & 273.8 MB & 209 KB & $\sim$ ~274 MB (28.3\%) & 71.7\% \\
    \bottomrule
  \end{tabular}
  \label{tab:table}
\end{table}

\section{Discussion}
Revisiting the aims set out in the introduction, we have been able to create a pipeline protocol that is robust for the de-identification, sequestering of relevant patient information, ROI identification and file compression of ultrasound images. Previous work has sought to train CNNs to perform detection and therefore removal of solely PHI related information contained within the image, with success ranging from 65-89\% \cite{monteiro2017identification} \cite{huang2009privacy} \cite{lien2011open}. Some of these techniques are operating system (OS) specific or only available at a cost \cite{rodriguez2010open}. The PyLogik package addresses these problems. It ensures the removal of direct patient identifiers while providing text file format conversion to .csv file output, correct ROI identification and information compression. In addition, the protocol is OS-agnostic and free of charge for researchers. By simplifying the deep learning problem– (removal/sequestering of all text versus specific text) we overcome the risk of distinguishing characters such as "Bg", from "B9", "B1” from "Bl" or "B0 or "Bo” and placed the necessary context specific filtering back on individual sites to perform on their .csv files. Because PyLogik can run on any OS and is free to download, it can run on servers behind institutional firewalls. By extracting and subsequently masking all text, the .csv files output by the pipeline allows end users to query, include, or destroy information for their specific uses.

Our strategy also facilitates better multimodal integration of data information. For example, on echocardiographic images, the heart rate is often displayed as text in each view; in PyLogik this information is retained, made available to end users, and is thus available for use during information fusion (early, joint, and late) in algorithm development \cite{kline2022multimodal}. Images are saved and output as JPEG stacks to decrease the number of specialized libraries and coding platforms needed to re-import the images for processing \cite{liu2007medical}. By truncating the image to only contain the ROI, we retain only salient information, thus facilitating compression on secondary non-PACs servers. 

In addition to providing an efficient de-identification and image cleaning protocol to facilitate leveraging ultrasound images on aggregate for algorithm development, our proposed method offers a $\sim$72\% compression in comparison to the original DICOM files. Not only does this have implications for long-term storage of these large files, but it allows for substantially increased short-term storage when used for analyses. This package is designed to be modular, with a separate class for those seeking solely the de-identification procedure partition of the pipeline having been created. For example, other functions outlined above could be used on ultrasound data if the user is looking only for de-identification (deid.deid). The same part of the pipeline processing may be easily extended to other imaging modalities such as magnetic resonance imaging (MRI), computed tomography (CT), and other radiographs. For example, if a user was seeking de-identification and maximal compression from MRI, x-ray or CT data, they may consider (deid.deid\_one) where patient image information is represented by a singular area/volume. This is shown in Figure 6 below.  

\begin{figure}[h!]
  \centering
  \includegraphics[scale=0.6]{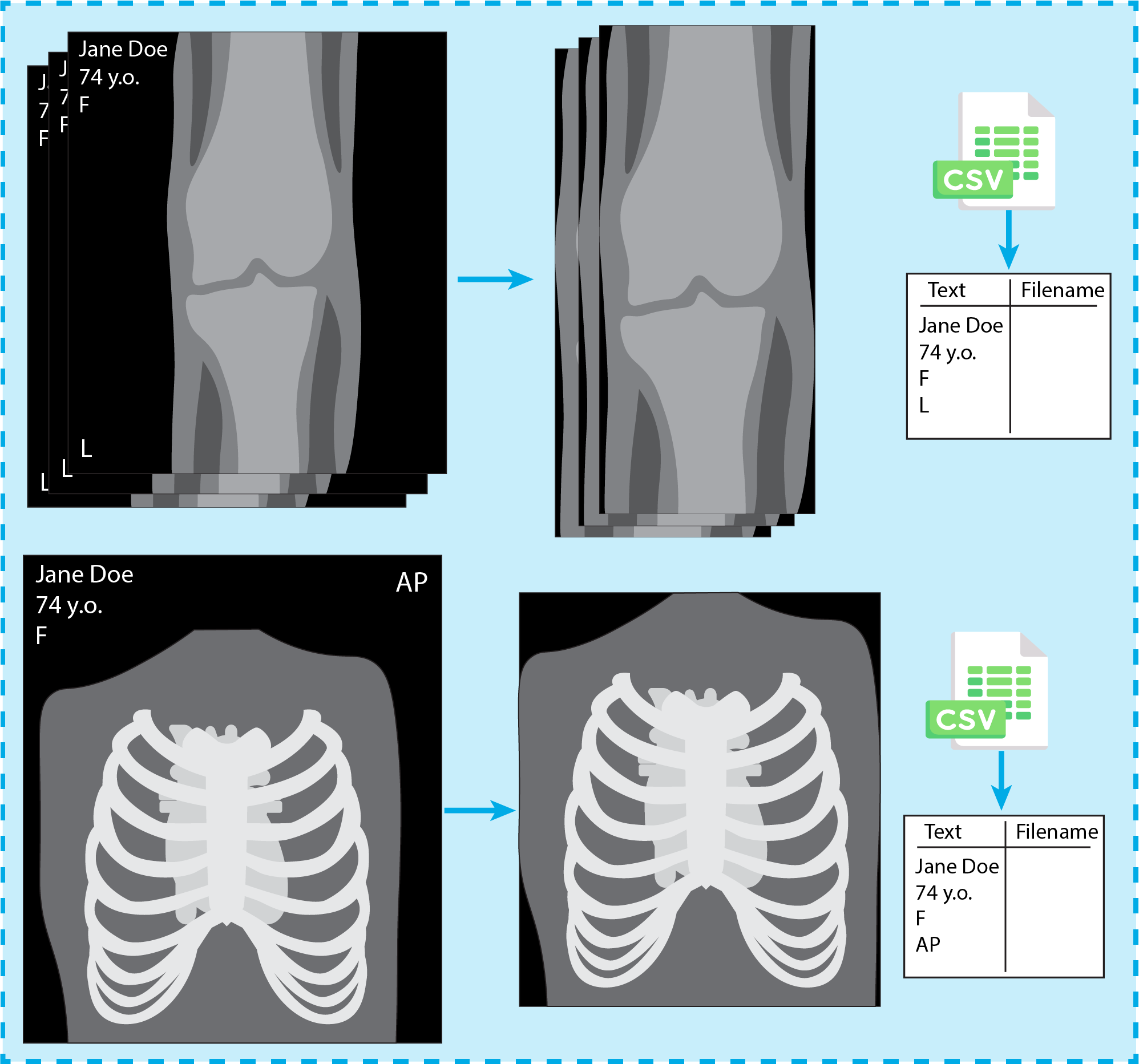}
  \caption{Demonstration of de-identification and compression for other imaging modalities (deid.deid\_one)}
  \label{fig:fig2} 
\end{figure}

Limitations to this work include that while automated data cleaning is desirable, rarely is an automated de-identification effort perfect. The risk of PHI leakage is of utmost concern due to legal and ethical ramifications. We urge the research community to test the protocol on their respective systems for ultrasound and image de-identification. Future work includes updates to the software package and incorporating feedback to make it more generalizable as adoption grows.

\newpage

\section{Conclusion}
This study showcases PyLogik, a python-based library for ultrasound de-identification with additional image processing features including DICE score visualization and value selection tools. It has been made readily available for ease of public use and dissemination. The implications of this efficient method for image data cleaning for use in a machine-learning pipeline are 1) deidentification when sharing data across centers to leverage inferences from ’big data’ 2) cleaning image volumes for algorithmic assessment, and 3) reduction of files size by an average fraction of 72\%. PyLogik is available at \url{https://pypi.org/project/pylogik/}.

\section*{Acknowledgments}
This work was supported by a grant from the National Institutes of Health (NIH; U54 HL160273 to SJS and YL), which includes a HeartShare Research Skills Fellowship Program, which provided mentorship, guidance, and funding to AK. VA is supported by a Pfizer Transthyretin Amyloidosis Cardiomyopathy Fellowship Program grant awarded to SJS. This work was also supported by additional NIH grants (R01 HL107577, R01 HL127028, R01 HL140731, and R01 HL149423) to SJS.

\section*{Contributions}
AK conceived the project and methodology, created the coding library, figure generation and manuscript writing. VA performed expert annotation of the echocardiograms, and proofreading/content curation of the final manuscript. SJS and YL provided mentorship, proofreading, and content curation of the final manuscript.

\section*{Disclosures}
SJS reports receiving research and educational grants from Pfizer; a research grant from Corvia Medical; and consulting fees from Abbott, Actelion, AstraZeneca, Amgen, Aria CV, Axon Therapies, Bayer, Boehringer-Ingelheim, Boston Scientific, Bristol-Myers Squibb, Cardiora, Coridea, CVRx, Cyclerion, Cytokinetics, Edwards Lifesciences, Eidos, Eisai, Imara, Impulse Dynamics, GSK, Intellia, Ionis, Ironwood, Lilly, Merck, MyoKardia, Novartis, Novo Nordisk, Pfizer, Prothena, Regeneron, Rivus, Sanofi, Sardocor, Shifamed, Tenax, Tenaya, Ultromics, and United Therapeutics. All other authors report no conflicts of interest. 

\bibliographystyle{unsrt}  
\bibliography{references}

\end{document}